# A diffuse-interface model for smoothed particle hydrodynamics

Zhijie Xu[1*], Paul Meakin[1,2,3] and Alexandre Tartakovsky[4]

1. Idaho National Laboratory Center for Advanced Modeling and Simulation.

2. Physics of Geological Processes, University of Oslo.

3. Multiphase Flow Assurance Innovation Center, Institute for Energy Technology, Kjeller.

4. Computational Mathematics Technical Group, Computational & Information Sciences Directorate, Pacific Northwest National Laboratory.

## Abstract

Diffuse-interface theory provides a foundation for the modeling and simulation of microstructure evolution in a very wide range of materials, and for the tracking/capturing of dynamic interfaces between different materials on larger scales. Smoothed particle hydrodynamics (SPH) is also widely used to simulate fluids and solids that are subjected to large deformations and have complex dynamic boundaries and/or interfaces, but no explicit interface tracking/capturing is required, even when topological changes such as fragmentation and coalescence occur, because of its Lagrangian particle nature. Here we developed an SPH model for single-component two-phase fluids that is based on diffuse-interface theory. In the model, the interface has a finite thickness and a surface tension that depend on the coefficient, $k$, of the gradient contribution to the Helmholtz free energy functional and the density dependent homogeneous free energy. In this model, there is no need to locate the surface (or interface) or to compute the curvature at and near the interface. One- and two-dimensional SPH simulations were used to validate the model.

---

* zhijie.xu@inl.gov

## I. INTRODUCTION

In the study of multi-component and single component multi-phase fluids, the geometry and properties of the interface(s) separating different phases is often the focus of attention. The classical singular (sharp) interface model developed by Young and Laplace in the early 1800's assumes that the interface is a surface of zero thickness. Within this paradigm, the free energy includes an excess surface free energy, which is proportional to the area of the interface, and this leads to the concept of surface tension. This sharp interface model has been successfully employed in a wide range of applications. However, a zero interface thickness and the associated discontinuous physical quantities across the interface, such as density and pressure, are non-physical.

The theory of diffuse (non-zero thickness) interfaces, originally developed by van der Waals [1] and refined by Cahn and Hilliard [2, 3], is based on the idea that a rapid but smooth transition between two adjacent, essentially homogeneous, bulk phases takes place across a thin interface zone. For many applications, in which important characteristic length scales are much larger than the interface width (typically on the order of 1 *nm*) the sharp interface description works very well. However, diffuse-interface theory provides an alternative but more realistic description of interfacial phenomenon, and it has been widely applied to physical processes with associated length scales that are comparable to the interfacial width, such as the critical point phenomenon [4], the motion of fluid-fluid-solid contact lines along a solid surfaces [5], solidification physics [6, 7], nucleation theory [8, 9], vapor condensation [10, 11] and many applications involving interfaces that undergo large deformations and/or topological changes (see [12] for an extensive review of the development of diffuse-interface theory and its application to single-component and binary fluids). Coupling of diffuses-interface theory with the Navier-Stokes equations provides a continuum (phase-field) approach to multiphase fluid flow and interface dynamics [13].

Most simulations of free surface flows and multiphase flows are performed using grid-based methods to solve continuum fluid dynamics equations, such as the Navier Stokes equations. However, smoothed particle hydrodynamics is quite widely used to simulate the behavior of multiphase materials subjected to large deformations. Despite its relatively low computational efficiency relative to grid-based computational dynamics, it is well suited to research applications because it allows fluid dynamics to be coupled with other physics, in a transparent manner, with relatively little code development effort. In this paper we developed a smoothed particle hydrodynamics (SPH) model that incorporates diffuse-interface theory to simulate multiphase fluid dynamics. SPH, a Lagrangian particle approach, was originally introduced by Lucy [14] and Monaghan [15, 16] in 1970s for astrophysical fluid dynamics applications. SPH has the advantages of explicit mass and linear momentum conservation. Additionally, because of the Lagrangian particle nature of SPH, explicit interface tracking is not required and the difficulties associated with the application of grid-based continuum numerical methods to processes with complex dynamic interfaces and/or boundaries is circumvented. More recently, SPH has been extensively applied to a wide range of free surface fluid flows [17], fluid flow in fractured and porous media [18, 19], phase separation of van der Waals fluid mixtures [20], and the behavior of solid materials under extreme loading and deformation conditions [21]. The purpose of this paper is to introduce an SPH model for diffuse interfaces and describe critical applications of the model. This model is based on the well-established diffuse interface theory for surface tension, and it retains the advantages of rigorous mass and momentum conservation.

In smoothed particle hydrodynamics, an ensemble of SPH particles with individual masses $m_i$, positions, $\mathbf{x}_i$ and, velocities, $\mathbf{v}_i$, is used to represent flowing fluids and/or deforming solids. SPH particles can be regarded as moving thermodynamic subsystems [22] that carrying various field variables (such as the density, $\rho_i$, and internal energy, $u_i$), depending on the particular application. The basic foundation of SPH is the use of interpolation kernels or smoothing functions, *W*, to compute field variables and their spatial gradients at any location. In general, the SPH formulation of continuum equations is not unique. One of the commonly used SPH formulations of continuum hydrodynamics is [23, 24],

$$\rho_i = \sum_{j=1}^{n_i} m_j W\left(r_{ij}, h\right), \tag{1}$$

$$\frac{d\mathbf{v}_i}{dt} = \sum_{j=1}^{n_i} m_j \left\{ \left( \frac{\boldsymbol{\sigma}_i}{\rho_i^2} + \frac{\boldsymbol{\sigma}_j}{\rho_j^2} \right) \bullet \frac{\partial W}{\partial r_{ij}} \mathbf{e}_{ij} \right\}, \quad \mathbf{e}_{ij} = \mathbf{r}_{ij} / r_{ij}, \tag{2}$$

where $\boldsymbol{\sigma}_i$ is the total stress tensor at particle *i*, and

$$\frac{du_i}{dt} = \frac{1}{2} \sum_{j=1}^{n_i} m_j \left\{ \left( \mathbf{v}_i - \mathbf{v}_j \right) \bullet \left( \frac{\boldsymbol{\sigma}_i}{\rho_i^2} + \frac{\boldsymbol{\sigma}_j}{\rho_j^2} \right) \bullet \frac{\partial W}{\partial r_{ij}} \mathbf{e}_{ij} \right\}, \tag{3}$$

where $u_i$ is the internal energy associated with particle *i*. In these equations, $\mathbf{r}_{ij} = \mathbf{x}_i - \mathbf{x}_j$, where $\mathbf{x}_i$ is the position of particle *i*, $n_i$ is the number of neighbors of particle *i* (the number of particles within a distance of *h*, the range of the smoothing function, *W*). Eqs. (1)-(3) are the discretized SPH version of mass conservation, linear momentum conservation and energy conservation. In general, the viscosity and equation of state are temperature dependent, and an SPH heat conduction term must be included in the energy equation [24, 25]. However, we are interested in the behavior of multiphase fluids under essentially constant temperature conditions in the subsurface, and we use a barotropic equation of state and omit the heat equation, as is common practice in many SPH simulations. The B-spline,

$$W(v,h)=\frac{\alpha_D}{h^D}\begin{cases}\left(1-3/2v^2+3/4v^3\right) & 0\le v<1\\ 1/4\left(2-v\right)^3 & 1\le v<2\\ 0 & otherwise\end{cases}, \tag{4}$$

is a widely used interpolation kernel (smoothing function), where $v=r_{ij}/h$, D is the spatial dimension, and $\alpha_D$ is a constant that assures proper normalization of the smoothing function ($\alpha_D = 2/3, 10/7\pi, 1/\pi$ for D = 1, 2 and 3).

## II. SPH FORMULATION FOR THE DIFFUSE-INTERFACE MODEL

Following the Cahn-Hilliard approach [2], the modified free energy (including both the reversible part of the internal energy $u_i$ and the free energy gradient contribution) associated with the system of *N* SPH particles is the sum of the free energies associated with each particle. The modified free energy *F* is given by

$$F=\sum_{i=1}^{N}m_i\left(A_i\left(\rho_i,T_i\right)+k\left|\nabla\rho\right|_i^2/\rho_i\right), \tag{5}$$

where *N* is the total number of particles, $A_i$ is the free energy per unit mass corresponding to the bulk free energy density, which is a state function of the mass density, $\rho_i$, and the temperature $T_i$. In Eq. (5), *k* is the "gradient energy" coefficient, which is assumed to be constant for simplicity. The simple quadratic form for the contribution of the density gradients to the free energy has been used extensively in diffuse interface models, and it can be interpreted using statistical mechanics [26]. A relationship between the pressure and the free energy density for each particle can be based on the standard thermodynamic law $p = -(\partial F / \partial V)_T$ ($p$ is the pressure, $F$ is the Helmholtz free energy, and $V$ is the volume), or

$$p_i = \rho_i^2 \frac{\partial A_i}{\partial \rho_i}. \tag{6}$$

Furthermore, the SPH particle free energy density $A_i$ may be written as,

$$A_i = -\int p_i(\rho_i)\, d(1/\rho_i) + C, \tag{7}$$

where $C$ is an arbitrary integration constant. The particle bulk free energy density $A_i$ depends only on the equation of state, and it is a single valued function of the particle density, $\rho_i$. The force on each particle can be written as the gradient of the modified total free energy $F$,

$$\mathbf{f}_i = m_i d\mathbf{v}_i / dt = -\nabla_i F, \tag{8a}$$

and

$$\nabla_i F = \sum_{j=1}^{n_i} m_j \left( \frac{\partial A_j}{\partial \rho_j} - \frac{k}{\rho_j^2} |\nabla \rho|_j^2 \right) \nabla_i (\rho_j) + \sum_{j=1}^{n_i} m_j \frac{2k}{\rho_j} (\nabla \rho)_j \cdot \left( \nabla_i (\nabla \rho)_j \right), \tag{8b}$$

where $\nabla_i$ stands for the gradient with respect to the position of particle $i$, i.e. $\nabla_i = \partial / \partial \mathbf{x}_i$, and this expression conserves the total energy of the system. The density gradient at particle $i$ is given by

$$(\nabla \rho)_i = \sum_{j=1}^{n_i} m_j \frac{\partial W}{\partial r_{ij}} \mathbf{e}_{ij}, \text{ and } (\nabla \rho)_j = \sum_{k=1}^{n_j} m_k \frac{\partial W}{\partial r_{jk}} \mathbf{e}_{jk}. \tag{9}$$

The corresponding gradient with respect to the position of particle *i* is given by,

$$\nabla_i \left(\nabla\rho\right)_j = \frac{\partial\left(\nabla\rho\right)_j}{\partial \mathbf{x}_i} = -m_i\mathbf{S}_{ij}\text{, and } \nabla_i \left(\nabla\rho\right)_i = \frac{\partial\left(\nabla\rho\right)_i}{\partial \mathbf{x}_i} = \sum_{j=1}^{n_i(j\neq i)} m_j\mathbf{S}_{ij}\ , \tag{10}$$

where $\mathbf{S}_{ij}$ is a second order tensor given by,

$$\mathbf{S}_{ij} = \nabla_i\left(\frac{\partial W}{\partial r_{ij}}\mathbf{e}_{ij}\right) = \frac{\partial^2 W}{\partial r_{ij}^2}\mathbf{e}_{ij}\otimes\mathbf{e}_{ij} + \frac{1}{r_{ij}}\frac{\partial W}{\partial r_{ij}}\left(\mathbf{I}-\mathbf{e}_{ij}\otimes\mathbf{e}_{ij}\right). \tag{11}$$

Here $\otimes$ stands for the tensor product and $\mathbf{I}$ is the unit matrix. By substituting Eq. (9) and Eq. (10) into Eq. (8b), and making use of the Eq. (6), we obtain the SPH particle equation of motion ,

$$\frac{d\mathbf{v}_i}{dt} = \sum_{j=1}^{n_i(j\neq i)} m_j\left\{\left[\underbrace{-\left(\frac{p_i}{\rho_i^2}+\frac{p_j}{\rho_j^2}\right)}_{1}+\underbrace{\left(\frac{k\left|\nabla\rho\right|_i^2}{\rho_i^2}+\frac{k\left|\nabla\rho\right|_j^2}{\rho_j^2}\right)}_{2}\right]\frac{\partial W}{\partial r_{ij}}\mathbf{e}_{ij} - \left[\underbrace{\frac{2k\left(\nabla\rho\right)_i}{\rho_i}}_{3}-\underbrace{\frac{2k\left(\nabla\rho\right)_j}{\rho_j}}_{4}\right]\bullet\mathbf{S}_{ij}\right\} \tag{12}$$

In Eq. (12), term 1 (on the right-hand-side) is the thermodynamic pressure contribution, and terms 2, 3 and 4 are related to the gradient contribution. Comparison between Eq. (12) and Eq. (2) used in the traditional SPH model indicates that the bulk free energy density $A_i$ is from the reversible part of the total free energy due to the work done by the pressure. The symmetry of Eq. (12) over indices *i* and *j* also ensures conservation of the total momentum of the SPH particle system. The effect of viscosity (in addition to the intrinsic momentum diffusion contribution due to particle motion relative to the local continuum flow velocity) can be easily incorporated by adding the corresponding viscosity terms into Eq. (12). Finally, the modified SPH equation of motion and energy equation are:

$$\frac{d\mathbf{v}_i}{dt} = \sum_{j=1}^{n_i(j\neq i)} m_j \left\{ \left[ \left( \frac{\boldsymbol{\sigma}_i}{\rho_i^2} + \frac{\boldsymbol{\sigma}_j}{\rho_j^2} \right) + \left( \frac{k|\nabla\rho|_i^2}{\rho_i^2} + \frac{k|\nabla\rho|_j^2}{\rho_j^2} \right) \right] \frac{\partial W}{\partial r_{ij}} \mathbf{e}_{ij} - \left[ \frac{2k(\nabla\rho)_i}{\rho_i} - \frac{2k(\nabla\rho)_j}{\rho_j} \right] \bullet \mathbf{S}_{ij} \right\} \quad (13)$$

$$U_i = u_i + km_i |\nabla\rho|_i^2 / \rho_i \quad (14)$$

where $U_i$ is a modified total internal energy incorporating the gradient contribution. $u_i$ is the traditional SPH internal energy obtained from Eq. (3) (if the system is not isothermal and heat conduction is important, a corresponding temperature term should be included in Eq. (3)). Particle motion governed by these two equations, conserves the total energy, $U = \sum_{i=1}^{N} \left( U_i + \frac{1}{2} m_i |\mathbf{v}_i|^2 \right)$, of the SPH particle system. The SPH implementation of Eq. (13) and (14) assures conservation of the total energy *U* of the system and the correct density gradient contribution to the particle dynamics, and hence the correct fluid dynamics.

To establish the connections to the continuum formulation of the diffuse-interface model, we start from the free energy functional,

$$F = \int_\Omega \left( A(\rho) + k|\nabla\rho|^2/\rho \right) \rho dV \,, \quad (15)$$

which is an integral over the entire problem domain $\Omega$. The Lagrangian of the system is given by:

$$L = \int_\Omega \left( A(\rho) + k|\nabla\rho|^2/\rho \right) \rho dV - \mu \int_\Omega \rho dV \,, \quad (16)$$

where $\mu$ is the chemical potential (per unit mass), and it can be interpreted as the thermodynamic grand potential. Using the Euler-Lagrange equation, the Lagrangian is minimized when

$$\mu = A(\rho) + \rho\, \partial A / \partial\rho - 2k\nabla^2\rho \,. \quad (17)$$

It follows from the Noether's Theorem [27], that the capillary stress tensor that satisfies the mechanical equilibrium condition $\nabla \bullet \mathbf{T} = 0$ is given by

$$\mathbf{T} = \left( -p + 2k\rho\nabla^2\rho + k\left|\nabla\rho\right|^2 \right)\mathbf{I} - 2k\nabla\rho \otimes \nabla\rho \,. \tag{18}$$

The driving force is then given by,

$$\frac{d\mathbf{v}}{dt} = \frac{\nabla \bullet \mathbf{T}}{\rho} = -\nabla\mu = -\frac{\nabla p}{\rho} + 2k\nabla\left(\nabla^2\rho\right). \tag{19}$$

The first term on the right-hand-side of Eq. (19) is the hydrodynamic pressure gradient contribution and the second term reflects the effect of the density gradient contribution to the free energy. The connections between Eq. (19) and the corresponding SPH particle acceleration (Eq.(12)) can be established by SPH discretization of terms 1, 2, 3, and 4:

$$\frac{\nabla p}{\rho} \approx \sum_{j=1}^{n_i(j\neq i)} m_j \left\{ \left( \frac{p_i}{\rho_i^2} + \frac{p_j}{\rho_j^2} \right) \frac{\partial W}{\partial r_{ij}} \mathbf{e}_{ij} \right\}, \tag{20}$$

$$\frac{\nabla\left( k\left|\nabla\rho\right|^2 \right)}{\rho} \approx \sum_{j=1}^{n_i(j\neq i)} m_j \left\{ \left( \frac{k\left|\nabla\rho\right|_i^2}{\rho_i^2} + \frac{k\left|\nabla\rho\right|_j^2}{\rho_j^2} \right) \frac{\partial W}{\partial r_{ij}} \mathbf{e}_{ij} \right\}, \tag{21}$$

$$\frac{\nabla\left( k\left|\nabla\rho\right|^2 \right)}{\rho} \equiv \frac{2k\nabla\rho \cdot (\nabla\nabla\rho)}{\rho} \approx \frac{2k\left(\nabla\rho\right)_i}{\rho_i} \bullet \sum_{j=1}^{n_i(j\neq i)} m_j \mathbf{S}_{ij} \,, \tag{22}$$

and $$2k\nabla\left(\nabla^2\rho\right) \equiv 2k\left(\nabla\nabla\right) \bullet \nabla\rho \approx \sum_{j=1}^{n_i(j\neq i)} m_j \left\{ \frac{2k\left(\nabla\rho\right)_j}{\rho_j} \bullet \mathbf{S}_{ij} \right\}. \tag{23}$$

## III. RESULTS AND DISCUSSION

An equation of state is required to implement the SPH simulation, and the van der Waals's (vdW) equation of state [28] is a popular and convenient choice for phase behavior applications because its analytical form is motivated by simple molecular concepts [29] and it produces an equation of state that is similar to that of molecular liquids. The vdW equation of state, can be expressed in the form

$$p = \frac{\rho \bar{k}}{1 - \bar{b}\rho} - \bar{a}\rho^2 , \qquad (24)$$

where $p$ is the pressure, $\rho$ is the fluid density and the parameters are defined as $\bar{k} = k_B T / m$, $\bar{a} = a / m^2$, and $\bar{b} = b / m$. Here, $k_B$ is the Boltzmann constant and $m$ is the particle mass. The vdW parameter $a$ is a measure of the attractive force between particles and $b$ is an excluded volume resulting from the short range repulsive particle-particle interactions. The particle free energy density $A$ obtained according to Eq. (7) is

$$A = -\left\{ \bar{a}\rho + \bar{k} \ln\left( 1/\rho - \bar{b} \right) \right\} + C , \qquad (25)$$

where $C$ is an arbitrary integration constant. The SPH particle free energy density can be computed easily from Eq. (25), and hence the total free energy of the system can be computed from Eq. (5). A schematic plot of the vdW equation of state is presented in Fig. 1, together with the free energy density, $A$, which is shown on the same plot. The stable free energy density minimum at $\rho_1 = \rho_L$ corresponds to the liquid phase. The free energy density maximum at $\rho_2$ is an unstable stationary point. Both $\rho_1$ and $\rho_2$ correspond to zero pressure. The surface tension (the excess free energy per unit area of interface relative to the free energy density of a homogeneous system with the same mass and average density, but with no interfaces) is given by

$$\Gamma = 2k \int_{-\infty}^{+\infty} (d\rho / dx)^2 dx = 2k \int_{\rho_g}^{\rho_L} (d\rho / dx) d\rho , \qquad (26)$$

for a planar interface, where $x$ is the coordinate along the interface normal. The equilibrium equation obtained from Eq. (19) is:

$$\frac{dv}{dx} = \frac{dp}{dx} - 2k\rho\frac{d^3\rho}{dx^3} = 0 \quad \text{, or equivalently,} \quad \frac{d^2\Phi(\rho)}{d\rho^2} = \frac{1}{\rho}\frac{dp}{d\rho}, \tag{27}$$

where $\Phi(\rho) = k(d\rho/dx)^2$. It follows from Eq. (27) that $\Phi(\rho)$, which depends only on the equation of state, is given by

$$\Phi(\rho) = \int \frac{p(\rho)}{\rho} d\rho - \iint p(\rho) d\left(\frac{1}{\rho}\right) d\rho + C_1\rho + C_2, \tag{28}$$

where $C_1$ and $C_2$ are two integration constants. The surface tension is given by

$$\Gamma = \left(2\int_{\rho_g}^{\rho_L} \sqrt{\Phi(\rho)} d\rho\right)\sqrt{k} = C_3\sqrt{k}, \tag{29}$$

and

$$C_3 = 2\int_{\rho_g}^{\rho_L} \sqrt{\Phi(\rho)} d\rho, \tag{30}$$

which depends only on the equation of state. Since the total free energy of the system is explicitly expressed by Eqs. (5) and (7), the stable particle configuration can be efficiently obtained by determining the particle configuration (the particle positions) that minimize the total free energy *F*. A conjugate gradient algorithm was used to find a stable particle configuration at a free energy minimum. The 1-dimension simulation results shown in Fig. 2 support the scaling law $\Gamma = C_3\sqrt{k}$ with $C_3 \approx 0.36$, compared to the analytical value $C_3 = 0.32$ from Eq. (30) with the vdW parameters $\bar{a} = 2.2$, $\bar{b} = 0.5$, $\bar{k} = 1$, $m_i = 1$ and an SPH smoothing length of $h = 1.4$. One possible reason for the deviation from the analytical value is from the SPH particle discretization [30], which is inherent to the SPH methods. The particle discretization gives rise to the problem of particle deficiency near or on the boundary. The support domains of boundary particles intersect the boundary, result in the non-zero surface integral and contribute to the computation error.

Nugent and Posch [24] showed that standard SPH simulations performed using a vdW equation of state, with no gradient contribution, do not form stable liquid drops. However, they were able to generated stable drops by increasing the smoothing length of the attractive forces corresponding to the long-range cohesive interactions in vdW fluids by a factor of two. However, this comes at the price of a substantially increased computational effort. In other SPH simulations, pair-wise particle-particle interaction with short range repulsive and relatively long range attractive interactions were added to SPH simulations with an ideal gas equation of state to bring about phase separation and generate surface tension [18]. Here we show that the diffuse-interface SPH model can be used to simulate single-component multiphase fluids with stable interfaces. Fig. 3 shows that when the energy of 225 particles, that initially formed a regular square array, was minimized a stable circular liquid drop formed due to the surface tension generated by the gradient contribution to the free energy gradient. It is quite likely that there are many configurations that correspond to local energy minima, and both simulation of the SPH particle equation of motion and energy minimization may result in a local minimum configuration. In problems of this type, involving a large number of particles, it is impossible to find the global minimum (the minimization is NP complete in the number of particles), but the local minima are all very similar to the global minimum. Fig.4 plots the density profile of the equilibrium circular drop along the radial direction. The pressure in circular drops can be computed from the Virial equation [31], and the surface tension can then be determined using the Young-Laplace equation, $\Gamma = p_l R$, where $p_l$ is the pressure in the interior of the drop and $R$ is the equilibrium radius of the liquid drop. Several liquid drops with various radii from 6 to 11 (in the unit of $h$) were obtained via energy minimization. Figure 5 shows the variation of the pressure $p_l$ with the droplet radius and a surface tension of $\Gamma = 0.41$ was obtained from the Young-Laplace equation.

To validate the diffuse-interface SPH model, two-dimensional SPH dynamics simulations of the small-amplitude oscillations of an inviscid vdW drop were performed. At each time step, the particle densities were calculated using Eq. (1), and the corresponding particle pressures were computed from the equation of state Eq. (24). The particle acceleration and internal energy $U_i$ were then computed from Eqs. (13) and (14), and the new particle velocities and positions were updated by time integration using the explicit "velocity Verlet" algorithm with a time step $\Delta t = 0.001$. The surface tension can be computed using [24],

$$\tau = 2\pi\sqrt{\frac{R^3\rho}{6\Gamma}}\,, \tag{31}$$

where $\tau$ is the period of oscillation and $\rho$ is the average fluid density. An equilibrium circular drop (225 particles prepared by free energy minimization) was first deformed into an elliptic shape using a density-conserving affine transformation (c.f. [24]), and then the ensemble of SPH particles was allowed to oscillate by running a dynamics simulation with zero viscosity. Fig. 6 shows the time evolution of the diameters ($R_x$ and $R_y$) measured along the principal axes of the drop. From the measured period, $\tau = 94$, a surface tension of $\Gamma = 0.409$ was obtained, in good agreement with the result from the Young-Laplace equation. There is a small momentum diffusion viscosity due to the particle nature of the model, which dampens oscillations, but does not change the oscillation frequency. Compared to the standard SPH method, several additional terms ($\nabla\rho$, $|\nabla\rho|^2$, and $\mathbf{S}_{ij}$) need to be computed in the SPH particle equation of motion (Eq. (12)). Since those terms can be computed within the loop of find neighbors for the particular SPH particle, there is only a slight increase of the computation time in comparison with the standard SPH method.

## IV. CONCLUSION

We show that diffuse-interface theory can be combined with smoothed particle hydrodynamics to provide a simple model for multiphase fluid dynamics. One- and two-dimensional simulations based on the diffuse interface SPH model were shown to be in good agreement with analytical results obtained from the vdW Cahn-Hilliard model. The model provides a simple easily implemented way of simulating liquid-vapor systems without the need of interface tracking. We expect that extension to multicomponent multiphase systems and three-dimensional systems will be straightforward.

## ACKNOWLEDGMENTS

This work was supported by the U.S. Department of Energy, Office of Science Scientific Discovery through Advanced Computing Program. The Idaho National Laboratory is operated for the U.S. Department of Energy by the Battelle Energy Alliance under Contract DE-AC07-05ID14517 and the Pacific Northwest National Laboratory is operated for the U.S. Department of Energy by Battelle under Contract DE-AC06-76RL01830.

FIG.. 1. A schematic plot of the vdW equation of state (pressure p vs. mass density $\rho$ Eq. (24)) and the corresponding free energy density *A* (Eq. (25)). The free energy minimum and maximum at zero pressure *p* are denoted by $\rho_1$ and $\rho_2$, respectively.

FIG.. 2. Plot of surface tension $\Gamma$ as a function of the energy gradient constant *k* from 1-D simulations. The surface tension was computed from the excess surface free energy with various gradient constants *k*. The straight line (least squares fit to the data) is $\Gamma = 0.36\sqrt{k}$ .

FIG.. 3. The initial configuration with SPH particles on the nodes of a 15X15 square lattice and the final (minimum free energy) configuration.

FIG.. 4. The density profile along the radial direction of the equilibrium liquid drop in Fig. 3.

FIG.. 5. Dimensionless fluid pressure at liquid droplet center for droplets of various radii. A linear fit shows an equivalent surface tension of $\Gamma = 0.41$ from the Young-Laplace equation.

FIG.. 6. The time evolution of the principal radii of an elliptic vdW liquid drop during a small-amplitude oscillation simulation.

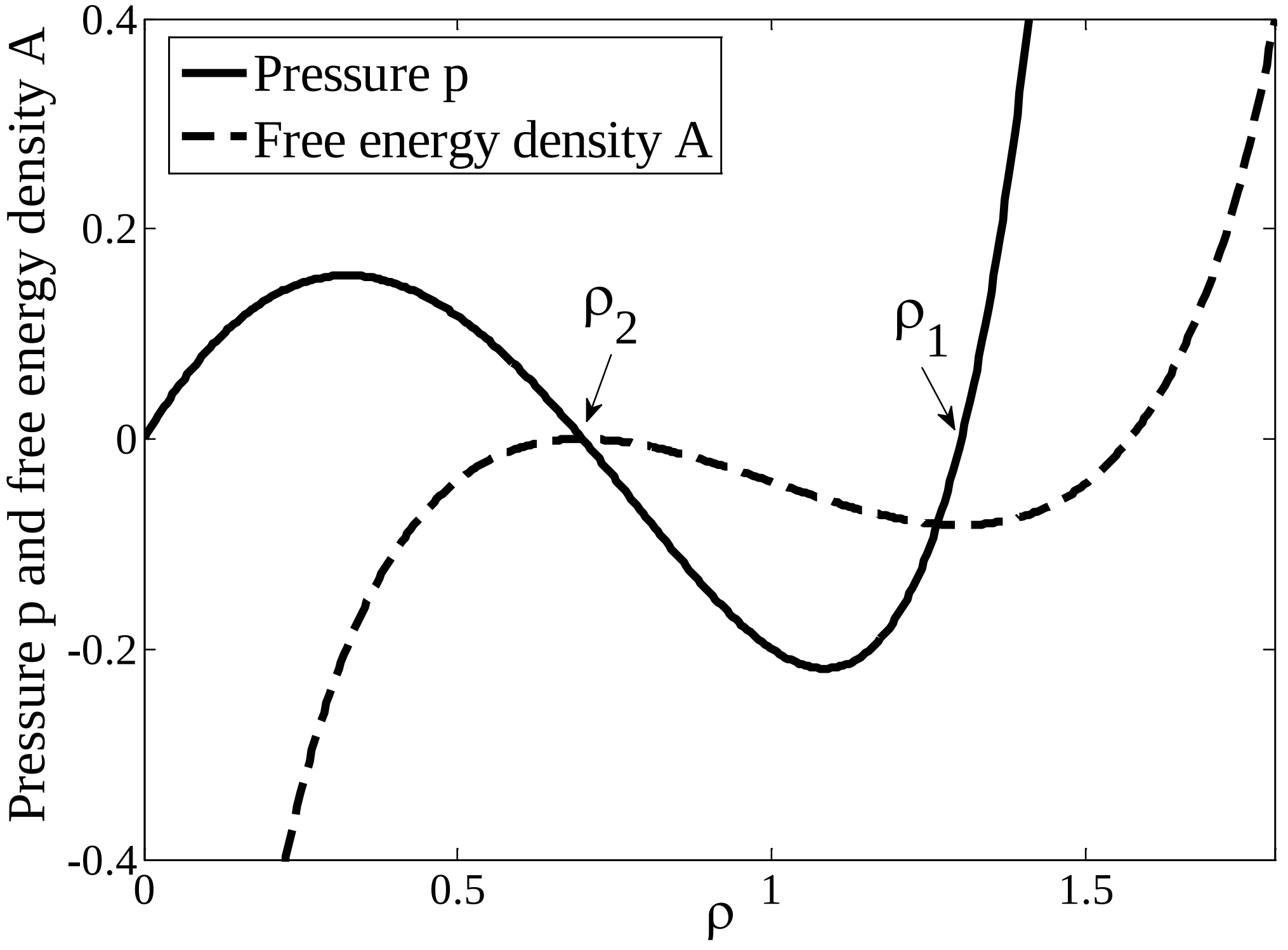

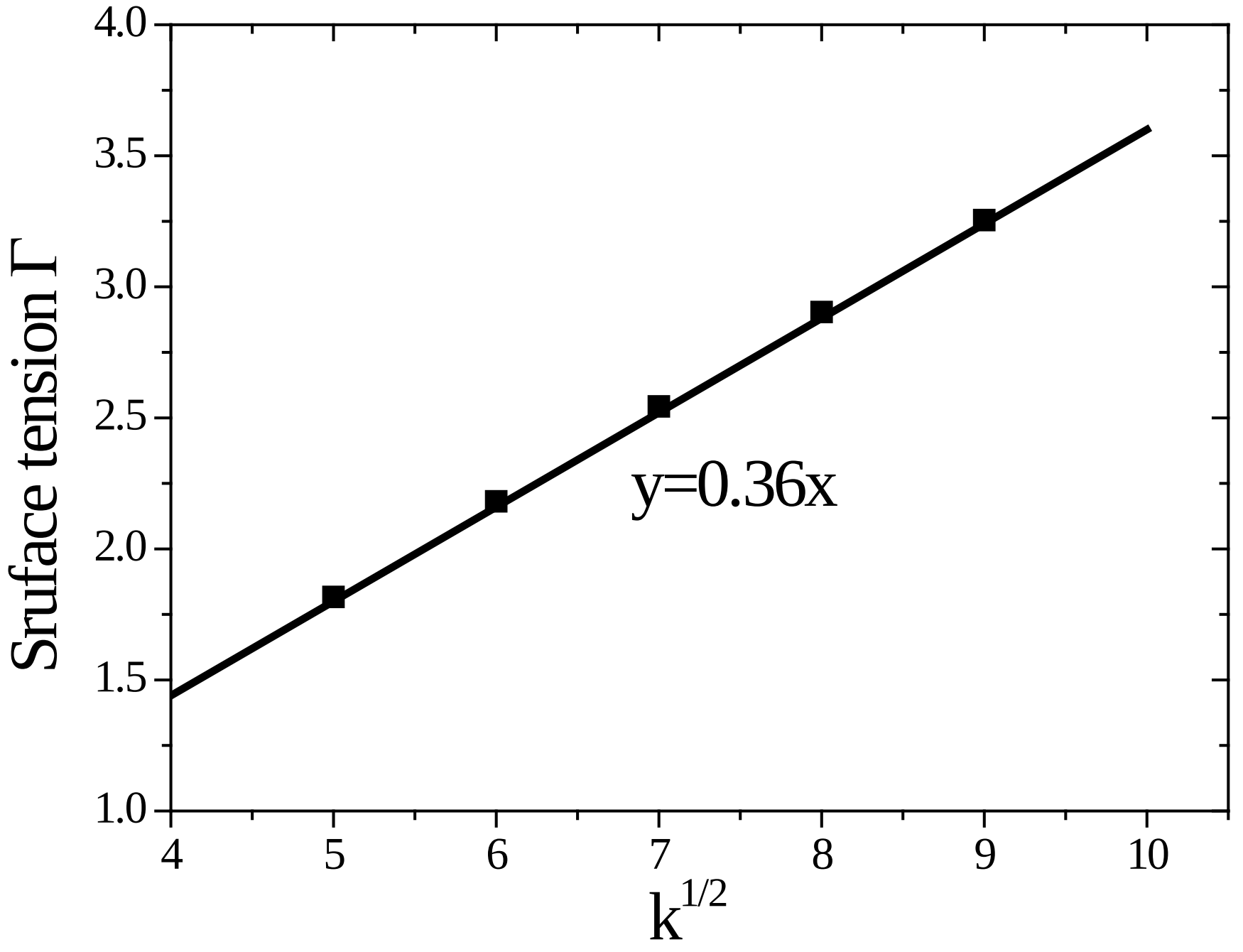

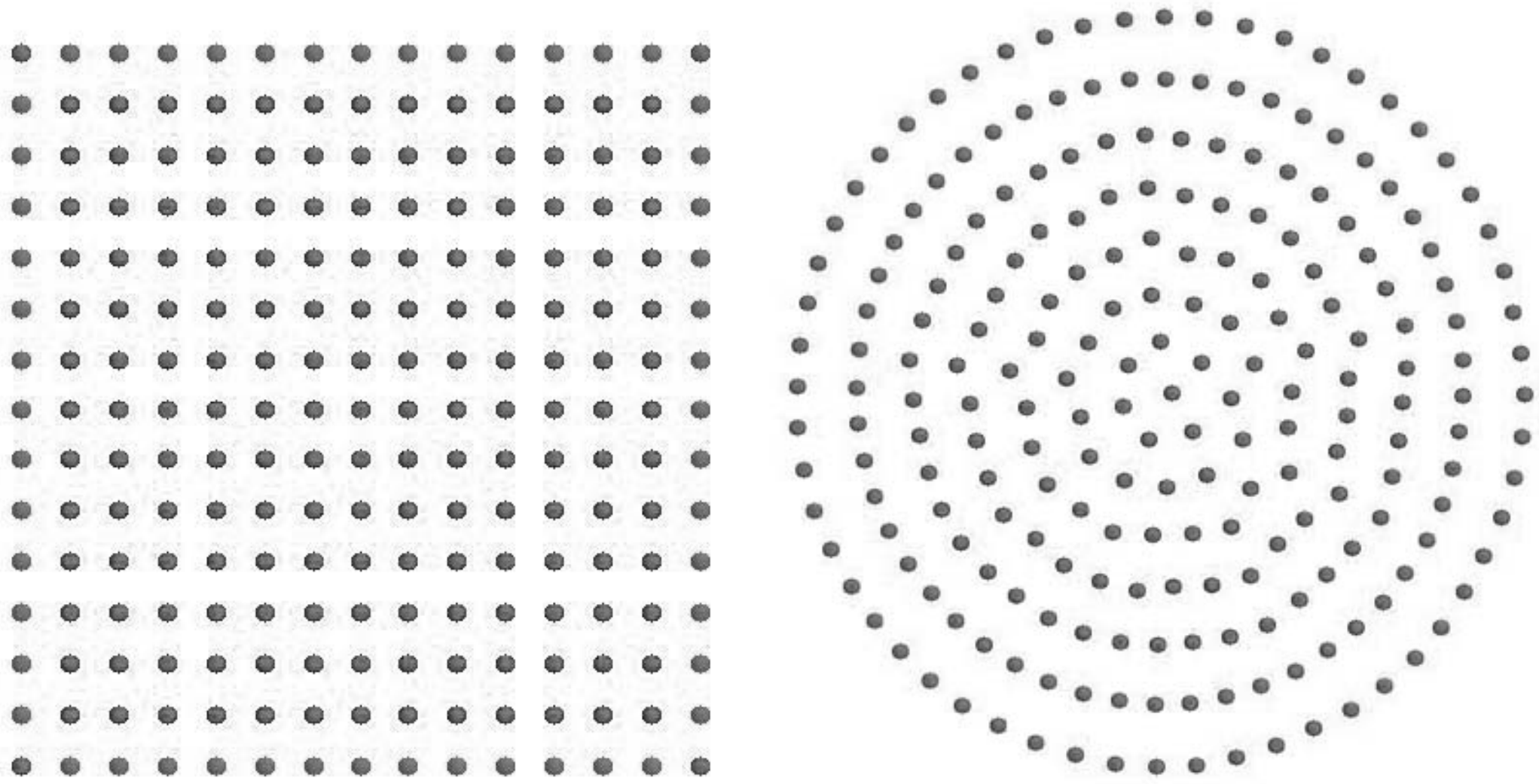

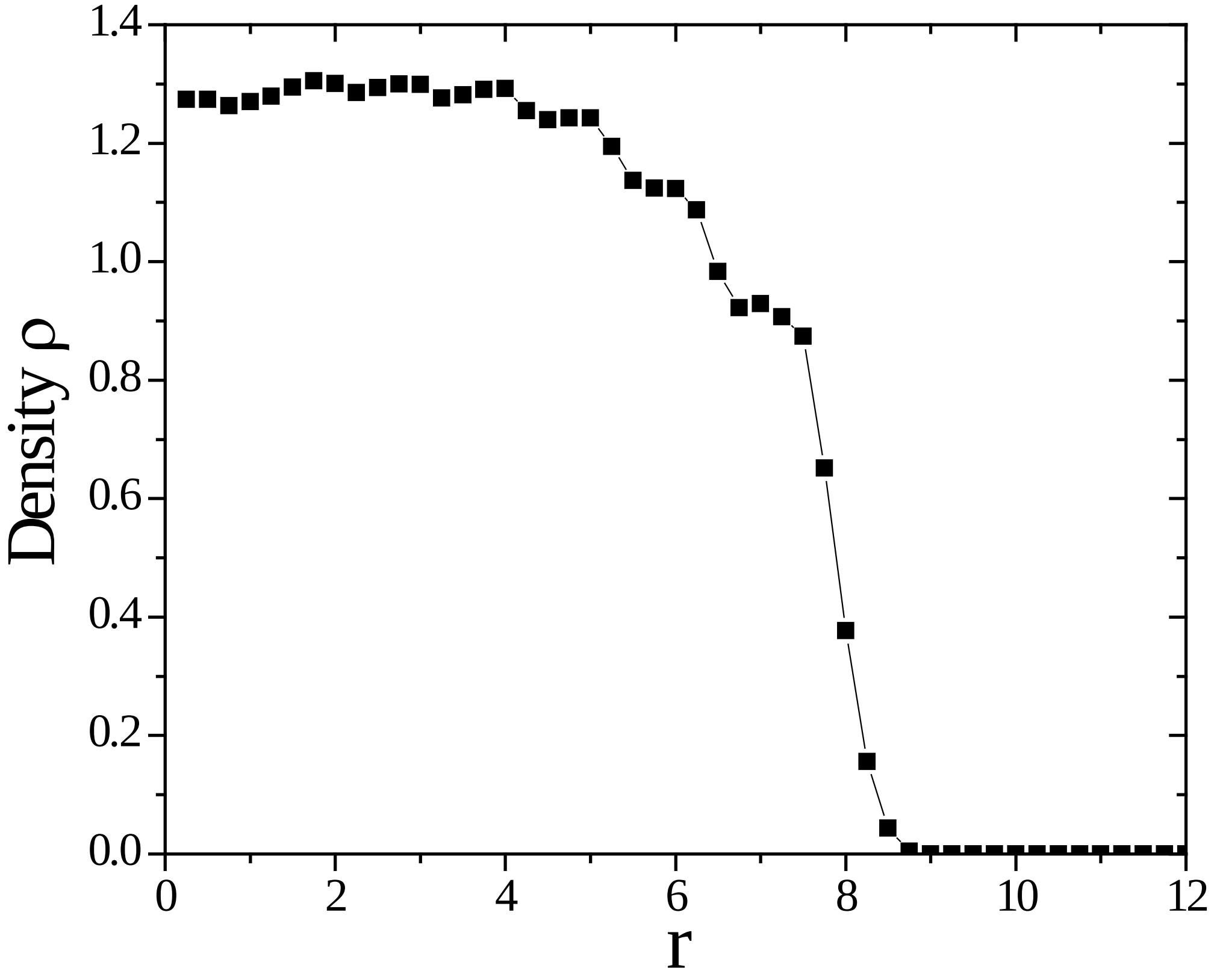

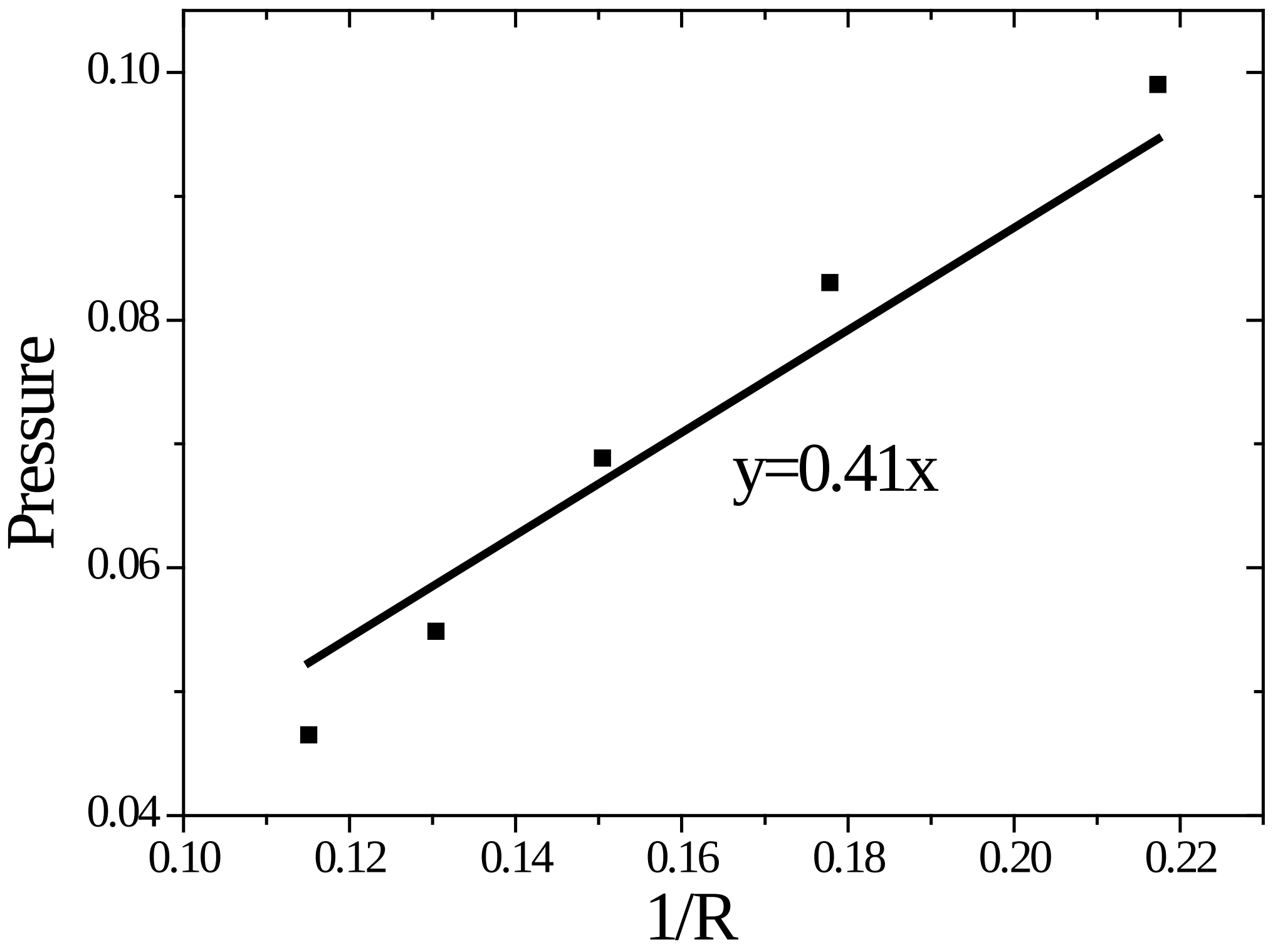

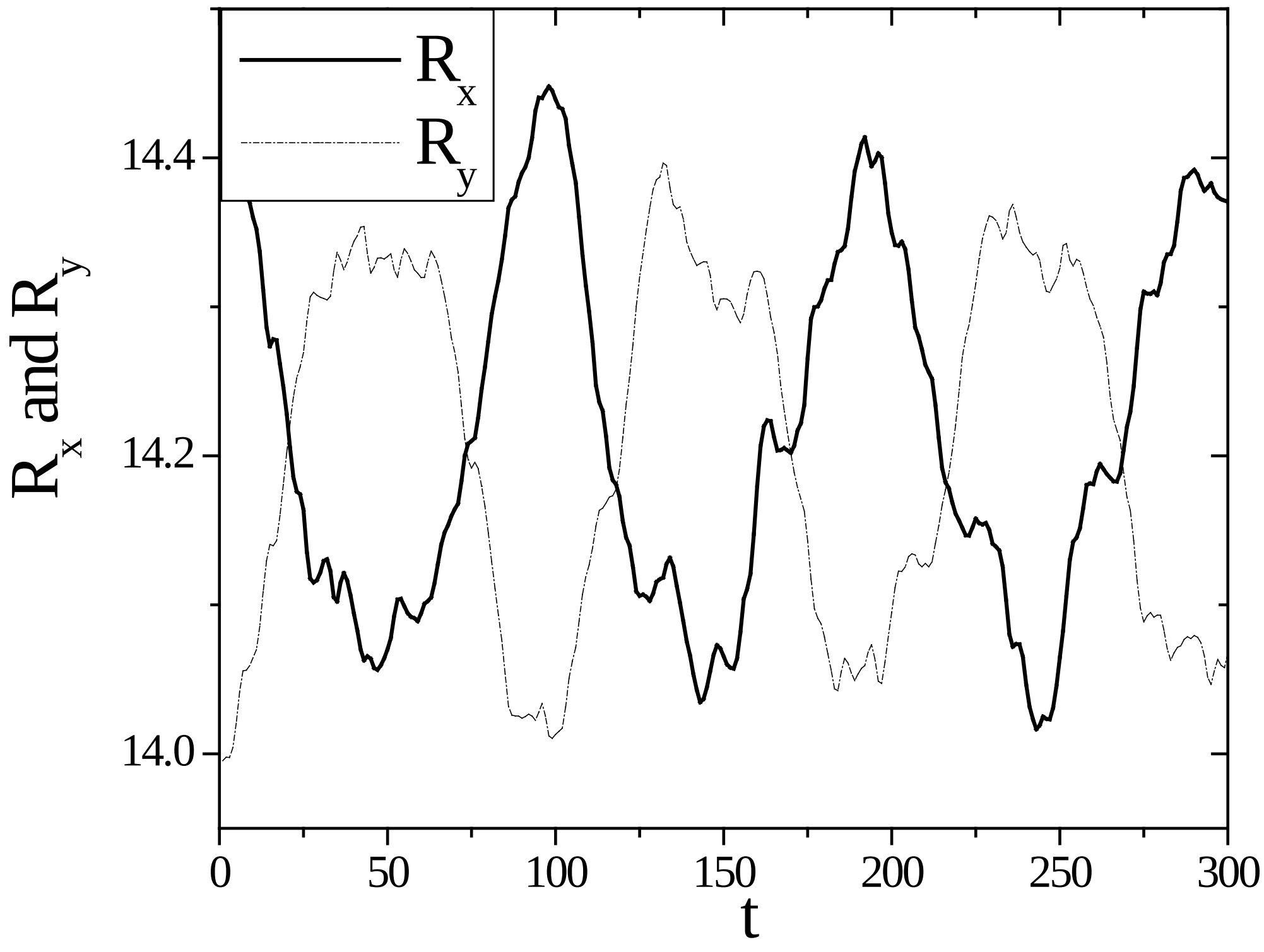